\documentclass[aps,twocolumn,showpacs]{revtex4}
\usepackage{graphicx}

\newcommand{\nc}{\newcommand}
\nc{\rnc}{\renewcommand}

\nc{\beq}{\begin{equation}}
\nc{\eeq}{\end{equation}}
\nc{\bea}{\begin{eqnarray}}
\nc{\eea}{\end{eqnarray}}
\nc{\ba}{\begin{array}}
\nc{\ea}{\end{array}}
\nc{\bpi}{\begin{picture}}
\nc{\epi}{\end{picture}}
\nc{\nn}{\nonumber}

\nc{\p}{\partial}
\nc{\f}[2]{\frac{#1}{#2}}
\nc{\od}{{\cal O}}
\nc{\ra}{\rightarrow}
\nc{\Rcal}{\cal R}
\nc{\uh}{\hat{u}}

\nc{\al}{\alpha}
\nc{\be}{\beta}
\nc{\de}{\delta}
\nc{\om}{\omega}
\nc{\ze}{\zeta}
\nc{\De}{\Delta}
\nc{\Si}{\Sigma}

\begin{document}

\bibliographystyle{apsrev}

\title{Shift of BEC Temperature of Homogenous Weakly Interacting Bose Gas}

\thanks{Talk presented at the 12th International Laser Physics Workshop,
LPHYS'03 (Hamburg, Germany, August 25-29, 2003).}

\author{Boris Kastening}
\affiliation{Institut f\"ur Theoretische Physik\\
Freie Universit\"at Berlin\\ Arnimallee 14\\ D-14195 Berlin\\ Germany\\
{\tt boris.kastening@physik.fu-berlin.de}}

\date{September 2003}

\begin{abstract}
We report on the computation of the shift of the Bose-Einstein condensation
temperature for a homogenous weakly interacting Bose gas in leading order
in the diluteness parameter $an^{1/3}$, where $a$ is the scattering length
and $n$ is the particle density.
The perturbative series, which is afflicted by infrared divergences,
is resummed by means of variational perturbation theory.
Using coefficients through seven loops, we arrive at
$\De T_c/T_c=1.27\pm0.11\,an^{1/3}$, which compares favorably with
recent Monte-Carlo data.
\end{abstract}

\maketitle

\section{Introduction}
The Bose-Einstein condensation (BEC) temperature $T_c$ of an ideal gas
of spin-0 Bosons is given by
\beq
\label{t0n0}
T_0=\f{2\pi}{m}\left[\f{n}{\ze(3/2)}\right]^{2/3},
\eeq
where $m$ is the mass of the bosons, $n$ their number density and we
work throughout in units where $k_B=\hbar=1$.
Infrared (IR) divergences prevent the perturbative evaluation of
the shift of the BEC temperature due to a small repulsive interaction,
parameterized by the corresponding $s$-wave scattering length $a$.
Physically, these IR divergences correspond to critical fluctuations
that change the universality class from Gaussian to that of a
three-dimensional scalar O(2)-symmetric field theory.

Although it is clear that $T_c$ changes by a small amount if the
dimensionless diluteness parameter $an^{1/3}$ is small (for small
momenta, $a$ is the only relevant parameter of the two-body interaction
potential), it was for a long time unclear, what the leading power in
terms of this parameter or the coefficient in front of it would be.
Many attempts on the problem have provided different powers and
coefficients \cite{history,Sto,other,GrCeLa,HoGrLa,HoKr,BaBlHoLaVa1,BaBlZi,%
WiIlKr,ArTo1,Al,SCPiRa1,KaPrSv,ArMo,HoBaBlLa,ArMoTo,BaBlHoLaVa2,SCPiRa2,%
KnPiRa,BrRa1,BrRa2,Kl1,Ka6loop,Ka7loop}.
Recently, however, it has been shown that the leading and next-to-leading
behavior is given by \cite{BaBlHoLaVa1,BaBlZi,HoBaBlLa}
\beq
\label{deltatexp}
\f{\De T_c}{T_0}=c_1an^{1/3}+[c_2'\ln(an^{1/3})+c_2''](an^{1/3})^2+\cdots
\eeq
In the work presented here, we determine $c_1$ numerically
(while $c_2'$ is known exactly, $c_2''$ may be determined by a
non-perturbative calculation analogous to the one presented here for
determining $c_1$;
for the exact value of $c_2'$ and a MC estimate of $c_2''$, see
Ref.\ \cite{ArMoTo}).
Results for $c_1$ in the literature range from $-0.93$ \cite{WiIlKr}
to $4.7$ \cite{Sto}, see Refs.\ \cite{ArMo,BaBlHoLaVa2} and the review
\cite{An}, also Fig.\ \ref{fig3} below.
In \cite{BaBlHoLaVa1} it was shown that $c_1$ is exclusively generated
by infrared fluctuations and that consequently three-dimensional field
theory is sufficient for the determination of $c_1$.
It appears that the most reliable results for $c_1$ before our work
have been obtained by Monte-Carlo (MC) simulations in three-dimensional
field theory \cite{KaPrSv,ArMo}.

Here we expand $\De T_c$ in a perturbative series, which is
subsequently resummed to obtain the physical non-perturbative limit.
While simple schemes such as Pad\'e approximants  may slowly converge
towards the correct value, our scheme of choice is Kleinert's field
theoretic variational perturbation theory (VPT, see \cite{Kl2,Kl3,Kl4}
and Chapters 5 and 19 of the textbooks \cite{pibook} and \cite{phi4book},
respectively; improving perturbation theory (PT) by a variational
principle goes back at least to \cite{Yu}), which has proven very useful
for the investigation of critical phenomena.

\section{Field Theory}
\label{fieldtheory}
\sloppy
The thermodynamic equilibrium properties of a homogenous gas of
spin-$0$ Bosons may be described in the grand canonical ensemble with
the help of a non-relativistic $(3+1)$-dimensional field theory in
imaginary time $\tau$, given by the Euclidean action
\beq
S_{3+1}
\!=\!\!
\int_0^\be\!\!\!d\tau\!\!\int\!\! d^3x\bigg[\psi^*\!\left(\f{\p}{\p\tau}
{-}\f{1}{2m}\nabla^2{-}\mu\right)\psi
+\f{2\pi a}{m}(\psi^*\psi)^2\bigg],
\eeq
where $\mu$ is the chemical potential.
This assumes that the momenta of the particles are small enough
so that their interaction potential is well described by just one
parameter, the $s$-wave scattering length $a$.
It also assumes that three- and more particle interactions are rare,
i.e.\ that the gas is dilute.
It turns out that the leading perturbative contribution to $\De T_c$ is
$\propto a^2$.
In consequence, the leading contribution to $\De T_c$ arises exclusively
from terms that are infrared divergent in the framework of PT.
This is why, for the determination of $c_1$, we can set all Matsubara
frequencies to zero or, equivalently, work with a three-dimensional field
theory \cite{BaBlHoLaVa1}.
Denoting the zero-Matsubara modes by $\psi_0$, we define the fields
and parameters of this theory by
$\psi_0=\sqrt{mT}(\phi_1+i\phi_2)$, $r_{\rm bare}=-2m\mu$, $u=48\pi amT$,
and obtain the three-dimensional Euclidean action
\beq
\label{s3}
S_3=\int d^3x\left[\f{1}{2}|\nabla\phi|^2
+\f{r_{\rm bare}}{2}\phi^2+\f{u}{24}(\phi^2)^2\right].
\eeq
For the determination of quantities that are not governed by the
leading IR divergences, one may still use a three-dimensional field
theory which, however, is obtained by a more complicated matching
procedure, which introduces corrections to the above relations of
the parameters of the $(3+1)$-dimensional and the three-dimensional
theory and also necessitates the inclusion of more interaction terms.
This matching is described in detail in \cite{ArMoTo}.

From the ideal-gas result (\ref{t0n0}) one obtains, in leading order,
the relation \cite{BaBlHoLaVa1}
\beq
\label{detc}
\f{\De T_c}{T_0}=-\f{2}{3}\f{\De n}{n},
\eeq
where $\De T_c$ is the shift of the condensation temperature for
fixed $n$ and $\De n$ is the shift of the critical particle density
for fixed condensation temperature.
In our field-theoretic setup, $\De n$ is given by
\beq
\label{deltanc}
\De n=\De\langle\psi^*\psi\rangle=mT\De\langle\phi^2\rangle.
\eeq
Combining (\ref{t0n0}), (\ref{deltatexp}), (\ref{detc}) and (\ref{deltanc}),
$c_1$ is given by
\beq
\label{c1def}
c_1=\al\left.\f{\De\langle\phi^2\rangle}{Nu}\right|_{\rm crit.}
\eeq
with $\al=-256\pi^3/[\ze(3/2)]^{4/3}\approx-2206.19$ and the remaining
task is to compute the critical limit of
\beq
\label{deltaphi2}
\f{\De\langle\phi^2\rangle}{Nu}
=\f{1}{u}\int\f{d^3p}{(2\pi)^3}[G(p)-G_0(p)]
\equiv\f{1}{u}\left[
\rule[-10pt]{0pt}{26pt}
\bpi(22,0)(2,0)
\put(13,2){\circle{14}}
\put(13,2){\circle{16}}
\put(13,-5){\makebox(0,0){$\times$}}
\epi
-
\rule[-10pt]{0pt}{26pt}
\bpi(22,0)
\put(13,3){\circle{16}}
\put(13,-5){\makebox(0,0){$\times$}}
\epi
\right],
\eeq
where $G$ and $G_0$ are the interacting and the free propagator,
respectively.
In (\ref{c1def}) and (\ref{deltaphi2}) we have conveniently generalized
the model (\ref{s3}) to an O($N$) field theory, where
$\phi=(\phi_1,\ldots,\phi_N)$, $\phi^2\equiv\phi_a\phi_a$.
This allows to make contact with the exactly known large-$N$ result
for $c_1$ \cite{BaBlZi} and with the MC results for $N=1$ and $N=4$
\cite{Su}.

The theory (\ref{s3}) is superrenormalizable.
A convenient and popular renormalization scheme is given by defining
a renormalized parameter $r=r_{\rm bare}-\Si(0)$, where $\Si(p)$
is the self-energy.
Now all Feynman diagrams can be computed without regularization.
The critical limit is obtained by letting $r\ra0$, as can be seen from
the structure of the full propagator, which reads now
$G(p)=1/\{p^2+r-[\Si(p)-\Si(0)]\}$.
For the free propagator follows $G_0(p)=1/(p^2+r)$ \cite{BrRa1}.

In a scheme used in \cite{SCPiRa1,SCPiRa2,KnPiRa,Kl1,HaKl}, the free
propagator is taken always critical, $G_0(p)=1/p^2$.
This generates an unnatural non-zero one-loop contribution to
$\De\langle\phi^2\rangle$, even in the absence of interactions.
Although this contribution vanishes as $r\ra0$, it strongly
influences resummation and is responsible for the small value of
$c_1=0.92\pm0.13$ \cite{Kl1res} in Ref.\ \cite{Kl1}.

\section{Perturbative Series for \boldmath$c_1$}
\label{pt}
In PT, $\De\langle\phi^2\rangle/Nu$ may be written as a power series
in $u_r\equiv Nu/4\pi r^{1/2}$.
Defining a function $c_1(u_r)$ with $c_1=\lim_{u_r\ra\infty}c_1(u_r)$,
we have
\beq
\label{pertsum}
c_1(u_r)=\al\sum_{l=1}^\infty a_lu_r^{l-2}.
\eeq
The inclusion of $N$ in the definition of $u_r$ is motivated by the fact
that then the $a_l$ remain finite even in the limit $N\ra\infty$,
facilitating the comparison with the exactly known large-$N$ result
\cite{BaBlZi}.
The successful application of VPT to the large-$N$ limit has been
demonstrated in \cite{Ka6loop}.

The perturbative series can be represented by Feynman diagrams and
then becomes a loop expansion.
The subtraction of the zero-momentum part of the self-energy in the
full propagator has to be performed recursively for all subdiagrams.
We denote this procedure by an operator $\cal R$ \cite{Ka6loop,Ka7loop}.
The first non-zero perturbative coefficient arises at the three-loop
level, since the lowest momentum-dependent self-energy contribution
has two loops.
The perturbative coefficient at $L$ loops is given by
\beq
a_Lu_r^{L-2}=\f{1}{Nu}\sum_{k=1}^{n_L}{\cal R}D_{L-k},
\eeq
where $n_L$ is the number of contributing diagrams in $L$-loop order
and $D_{L-k}$ is the $k$-th $L$-loop diagram.
There are $n_L=0,0,1,1,5,12,56$ diagrams at one- through seven-loop
order (see \cite{recrel} for a convenient way of constructing all
necessary diagrams together with their weights) and the expansion for
$c_1(u_r)$ starts out as 
\setlength{\unitlength}{0.275mm}
\bea
\lefteqn{c_1(u_r)}
\nn\\
&=&
\f{\al}{Nu}\Bigg[
{\cal R}
\rule[-14pt]{0pt}{34pt}
\bpi(34,0)
\put(17,3){\circle{24}}
\put(17,3){\oval(24,8)}
\put(5,3){\circle*{2}}
\put(29,3){\circle*{2}}
\put(17,-9){\makebox(0,0){$\times$}}
\epi
+{\cal R}
\bpi(34,0)
\rule[-14pt]{0pt}{34pt}
\put(17,3){\circle{24}}
\put(6.6,-3){\line(1,0){20.8}}
\put(6.6,-3){\line(3,5){10.4}}
\put(27.4,-3){\line(-3,5){10.4}}
\put(6.6,-3){\circle*{2}}
\put(27.4,-3){\circle*{2}}
\put(17,15){\circle*{2}}
\put(17,-9){\makebox(0,0){$\times$}}
\epi
+{\cal R}
\bpi(34,0)
\put(17,3){\circle{24}}
\put(8.5,-5.5){\line(1,0){17}}
\put(8.5,-5.5){\line(0,1){17}}
\put(8.5,11.5){\line(1,0){17}}
\put(25.5,-5.5){\line(0,1){17}}
\put(8.5,-5.5){\circle*{2}}
\put(8.5,11.5){\circle*{2}}
\put(25.5,-5.5){\circle*{2}}
\put(25.5,11.5){\circle*{2}}
\put(17,-9){\makebox(0,0){$\times$}}
\epi
+{\cal R}
\bpi(50,0)
\put(13,3){\circle{16}}
\put(37,3){\circle{16}}
\put(25,11){\oval(24,16)[t]}
\put(25,-5){\oval(24,16)[b]}
\put(13,-5){\line(0,1){16}}
\put(37,-5){\line(0,1){16}}
\put(13,-5){\circle*{2}}
\put(13,11){\circle*{2}}
\put(37,-5){\circle*{2}}
\put(37,11){\circle*{2}}
\put(5,3){\makebox(0,0){$\times$}}
\epi
\nn\\
&&{}
+{\cal R}
\bpi(50,0)
\put(13,3){\circle{16}}
\put(37,3){\circle{16}}
\put(25,11){\oval(24,16)[t]}
\put(25,-5){\oval(24,16)[b]}
\put(13,-5){\line(0,1){16}}
\put(37,-5){\line(0,1){16}}
\put(13,-5){\circle*{2}}
\put(13,11){\circle*{2}}
\put(37,-5){\circle*{2}}
\put(37,11){\circle*{2}}
\put(25,-13){\makebox(0,0){$\times$}}
\epi
+{\cal R}
\bpi(58,0)
\put(13,3){\circle{16}}
\put(45,3){\circle{16}}
\put(53,-5){\line(0,1){16}}
\put(33,-5){\oval(40,16)[b]}
\put(33,11){\oval(40,16)[t]}
\put(13,3){\oval(48,16)[r]}
\put(53,3){\circle*{2}}
\put(37,3){\circle*{2}}
\put(13,-5){\circle*{2}}
\put(13,11){\circle*{2}}
\put(5,3){\makebox(0,0){$\times$}}
\epi
+{\cal R}
\bpi(58,0)
\put(13,3){\circle{16}}
\put(45,3){\circle{16}}
\put(53,-5){\line(0,1){16}}
\put(33,-5){\oval(40,16)[b]}
\put(33,11){\oval(40,16)[t]}
\put(13,3){\oval(48,16)[r]}
\put(53,3){\circle*{2}}
\put(37,3){\circle*{2}}
\put(13,-5){\circle*{2}}
\put(13,11){\circle*{2}}
\put(33,-13){\makebox(0,0){$\times$}}
\epi
+\cdots\Bigg].
\nn\\
\eea
E.g., the coefficient $a_3$ is given by
\bea
\label{a3}
\lefteqn{a_3u_r}
\nn\\
&=&
\f{1}{Nu}\,{\cal R}
\rule[-14pt]{0pt}{34pt}
\bpi(34,0)
\put(17,3){\circle{24}}
\put(17,3){\oval(24,8)}
\put(5,3){\circle*{2}}
\put(29,3){\circle*{2}}
\put(17,-9){\makebox(0,0){$\times$}}
\epi
\nn\\
&=&
\f{(N+2)u}{18}
\int\f{d^3k}{(2\pi)^3}\int\f{d^3p}{(2\pi)^3}
\int\f{d^3q}{(2\pi)^3}\f{1}{(k^2+r)^2}
\nn\\
&&\times
\left[\f{1}{(k+p)^2+r}-\f{1}{p^2+r}\right]\f{1}{[(p+q)^2+r](q^2+r)}
\nn\\
&=&
-\f{(1+\f{2}{N})\ln\f{4}{3}}{576\pi^2}u_r.
\eea
The results for diagrams through six loops in the renormalization
scheme employed here can be found in \cite{MutNi}.
The seven-loop diagrams have been computed in 1991 and used
for the determination of critical exponents for $N=0,1,2,3$ in \cite{MurNi}.
They were provided to the author by B.~Nickel \cite{Nipc} and may be
found in \cite{Ka7loop}.
The resulting perturbative coefficients for $N=2$ are
$a_3=-1.01209\times10^{-4}$,
$a_4=2.99626\times10^{-5}$, $a_5=-1.19872\times10^{-5}$,
$a_6=5.85519\times10^{-6}$, $a_7=-3.30467\times10^{-6}$.

\section{Resummation and Results}
\label{resummation}
For obtaining the $u_r\ra\infty$ limit of $c_1(u_r)$, the series
(\ref{pertsum}) needs to be resummed.
For critical fluctuations phenomena, VPT is an appropriate tool, 
since it is designed to accommodate the large-$u_r$ behavior of
physical quantities in the critical limit, given in our case by
\beq
\label{larggeur}
c_1(u_r)=\al\sum_{m=0}^\infty f_mu_r^{-m\om'}
\eeq
with an irrational exponent $\om'$.
The leading class of corrections of a physical quantity that remains
finite in the critical limit are integer powers of $t^{\om\nu}$
\cite{phi4book,We,ZiPeVi}, where $t\equiv(T-T_c)/T_c$,
$\nu$ is the critical exponent of the correlation length and
$\om=\be'(g^*)$ in a renormalization group approach.
In our renormalization scheme, the propagator obeys
$G(p=0)=1/r\propto t^{-\gamma}$.
Using the universal scaling relation $\gamma=\nu(2-\eta)$,
where $\eta$ is the anomalous dimension of the critical propagator,
i.e.\ $G(r=0)\propto1/p^{2-\eta}$ in the small-$p$ limit,
we see that we have to set
\beq
\label{omp}
\om'=\f{2\om}{2-\eta}
\eeq
in (\ref{larggeur}).

The ansatz (\ref{larggeur}) does not account for so-called confluent
singularities which cause the true large-$u_r$ expansion to also contain
other negative powers of $u_r$, which are subleading at least compared
to $u_r^{-\om'}$.
We can expect methods that can accommodate the leading behavior in
(\ref{larggeur}) correctly to converge faster to the true result than
methods having the wrong leading behavior such as e.g.\ Pad\'{e}
approximants or the linear $\de$ expansion (LDE), the latter being
used extensively for the current problem
\cite{SCPiRa1,SCPiRa2,KnPiRa,BrRa1,BrRa2}.
On the other hand, convergence will be slowed by the fact that we
do not make an ansatz reflecting the full power structure in $u_r$,
but the expansion (\ref{larggeur}) will try to mimic the neglected
subleading powers.

The alternating signs of the $a_L$ suggest that the perturbative series
for $c_1(u_r)$ is Borel summable.
In the context of critical phenomena, such series have been successfully
resummed using Kleinert's VPT.
Accurate critical exponents \cite{Kl3,Kl4,phi4book} and amplitude ratios
\cite{KlvdB} have been obtained.
For a truncated partial sum $\sum_{l=1}^La_lu_r^{l-2}$ of (\ref{pertsum}),
the method requires replacing
\bea
u_r^{l-2}
\!&\ra&\!
(t\uh)^{l-2}
\left\{1+t\left[\left(\f{\uh}{u_r}\right)^{\om'}-1\right]
\right\}^{-(l-2)/\om'}
\eea
(note that this is an identity for $t=1$),
reexpanding the resulting expression in $t$ through $t^{L-2}$, setting
$t=1$ and then optimizing in $\uh$, where optimizing is done in accordance
with the principle of minimal sensitivity \cite{Ste} and in practice means
finding appropriate stationary or turning points.
That is, we replace
\beq
\label{uuhat}
u_r^{l-2}
\ra
\uh^{l-2}\sum_{k=0}^{L-l}
\left(\ba{c}-(l-2)/\om'\\k\ea\right)
\left[\left(\f{\uh}{u_r}\right)^{\om'}-1\right]^k
\eeq
and optimize the resulting expression in $\uh$.
For $u_r\ra\infty$, we obtain the $L$-loop approximation of $f_0$,
\beq
\label{f0}
f_0^{(L)}={\rm opt}_{\uh}\left[\sum_{l=1}^La_l\uh^{l-2}\sum_{k=0}^{L-l}
\left(\ba{c}-(l-2)/\om'\\k\ea\right)(-1)^k\right].
\eeq
Results are only available starting at four loops, since two non-zero
perturbative coefficients are necessary for VPT to work.
E.g., in four-loop order the optimization yields
\beq
f_0^{(4)}\approx{\rm opt}_{\uh}
\left[a_3\left(1+\f{1}{\om'}\right)\uh+a_4\uh^2\right]
=-\f{a_3^2\left(1+\f{1}{\om'}\right)^2}{4a_4}
\eeq
as the best attempt at determining $f_0$ and thus $c_1=\al f_0$.

For $N=2$, we have $\om=0.79\pm0.01$, $\eta=0.037\pm0.003$
(see, {\em e.g.}, \cite{phi4book}) and thus, using (\ref{omp}),
$\om'=0.805\pm0.011$.
The four- through seven-loop results are
$c_1=0.948$, $1.062$, $1.126$, $1.161$.

$\om'$ may also be determined self-consistently \cite{Kl3,pibook,phi4book}.
Assuming a behavior of $c_1(u_r)$ as in (\ref{larggeur}), the quantity
$d\ln c_1(u_r)/d\ln u_r$ has an expansion of the same type with the same
$\om'$ as $c_1(u_r)$, but with a vanishing large-$u_r$ limit (i.e.\ its
large-$u_r$ expansion starts out with $f_0=0$).
$\om'$ is tuned such that the value VPT gives for
$d\ln c_1(u_r)/d\ln u_r$ is zero in a given loop order.
This $\om'$ is then used as an input for the determination of the
approximation of $c_1(u_r)$ at the same loop order.
The five- through seven-loop results are
$c_1=1.399$, $1.383$, $1.376$.

From the fact that the fixed-$\om'$ results monotonically rise with
the loop order, while the results with self-consistent $\om'$ fall,
we estimate that the true value lies inbetween and conclude from the
resummed results through seven loops that
\beq
\label{c1n2}
c_1=1.27\pm0.11.
\eeq

Our treatment for $N=2$ may easily be repeated for arbitrary $N$.
E.g., for $N=1$ one obtains
\beq
\label{c1n1}
c_1=1.07\pm0.10,
\eeq
while for $N=4$, the result is
\beq
\label{c1n4}
c_1=1.54\pm0.11.
\eeq

\section{Discussion}
\label{discussion}
\begin{figure}[h]
\bigskip
\begin{center}
\includegraphics[width=8cm,angle=0]{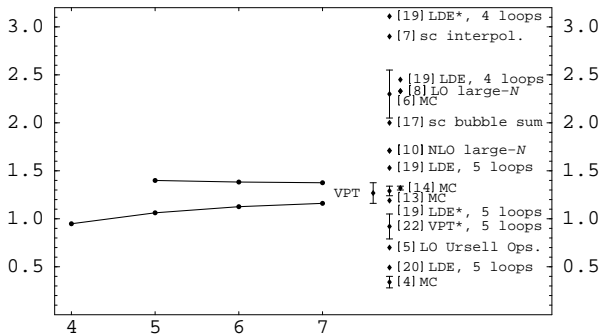}
\end{center}
\vspace{-15pt}
\caption{\label{fig3}
Comparison of $c_1$ from VPT as a function of the number of loops $L$ for
$N=2$ with most results from other sources.
Upper dots: self-consistent (sc) $\om'$.
Lower dots: $\om'=0.805$.
LDE$^\star$ indicates use of LDE with a resummation method for
accelerated convergence. 
The label VPT$^\star$ indicates the inclusion of a non-zero one-loop
term in Ref.\ \cite{Kl1}, which is absent in the present treatment,
labeled by VPT.}
\end{figure}

Many different results have been obtained for $c_1$;
see \cite{Ka7loop} and corresponding references therein for criticisms of
most of them.
A comparison of our results through seven loops with most other results
found in the literature is given in Fig.\ \ref{fig3}.
Our result (\ref{c1n2}) is in agreememt with the apparently most
reliable other sources available, namely the MC results from the
three-dimensional theory:
$c_1=1.29\pm0.05$ by Kashurnikov, Prokof'ev and Svistunov \cite{KaPrSv}
and $c_1=1.32\pm0.02$ by Arnold and Moore \cite{ArMo}.

MC results for $N=1$ and $N=4$ are also available.
X.~Sun has computed $\De\langle\phi^2\rangle_{\rm crit.}/u$ for these
cases \cite{Su} which translate to $c_1=1.09\pm0.09$ for $N=1$ and
$c_1=1.59\pm0.10$ for $N=4$.
Our values (\ref{c1n1}) and (\ref{c1n4}) agree well with these results.
This provides additional confidence that in fact both MC and our VPT
results are trustworthy estimates for $c_1$.

We would like to emphasize that the coefficient $c_1$ is of physical
relevance not only for the case of a strictly homogenous gas, but also
for a trap that is sufficiently wide to allow for critical fluctuations
\cite{ArTo1,ArTo2}.
If expressed as a function of the total number of particles $N_p$, the
leading effect on the shift of the condensation temperature is obtained
from mean field theory and only in second order the influcence of critical
fluctuations can be felt \cite{ArTo2,GiPiAt}.
In contrast, if expressed in terms of the central density $n$ of
particles, the leading effect is determined by critical fluctuations
as we have seen here.
It would therefore be desirable to know not only the total number of
particles $N_p$ (as in a recent experiment \cite{GeThRiHuBo}) but also the
central density of particles in experiments that measure the condensation
temperature.
In this context we would also like to stress that it is no contradiction
that the sign of the change of $T_c$ is different in both cases, since
different physical quantities $N_p$ and $n$ are kept fixed, respectively.
Let us turn this around:
For a given condensation temperature, a small repulsive interaction
causes the corresponding central density in a wide trap to be lowered
and the total number of particles to be raised compared to the ideal
gas case.

\section*{Acknowledgments}
The author is extremely grateful to B.~Nickel for providing the
results of his seven-loop calculations and thus facilitating the
extension of the VPT treatment through seven loops.
He thanks H.~Kleinert, P.~Arnold and E.~Braaten for valuable input.
Helpful communications with J.~Andersen, M.~Holzmann, F.~Lalo\"e,
D.~Murray, M.~Pinto and R.~Ramos are also acknowledged.


\begin{thebibliography}{99}
\bibitem{Kl1res}
The final result $c_1=1.14\pm0.11$ of \cite{Kl1} was obtained by also
applying VPT with self-consistent $\om'$ to scheme III of \cite{BrRa1}
(as first done in \cite{Ka6loop}) through five loops (albeit with the imprecise
numerical values for the integrals given in \cite{BrRa1})  and combining
the result with $c_1=0.92\pm0.13$.
\end{thebibliography}
\end{document}